\documentstyle[aps,prl,twocolumn,floats,psfig]{revtex}

\newcommand{\erfc}{ {\rm erfc} }
\newcommand{\const}{ {\rm const} }

\begin{document}

\bibliographystyle{prsty}

\title{ 
\begin{flushleft}
%{\small \em submitted to}\\
{\small 
PHYSICAL REVIEW E 
\hfill
VOLUME {\normalsize 60}, 
NUMBER {\normalsize 6} $\qquad$
\hfill 
DECEMBER,  {\normalsize 6499$-$6502} 
}
\end{flushleft}  
Thermally activated escape rates of uniaxial spin systems with transverse field:\\
Uniaxial crossovers
\vspace{-1mm}
}

\author{
D. A. Garanin \cite{e-gar}, 
}

\address{
Max-Planck-Institut f\"ur Physik komplexer Systeme, N\"othnitzer Strasse 38,
D-01187 Dresden, Germany\\ }

\author{
E. C. Kennedy\cite{e-ken}, D. S. F. Crothers
}

\address{
Dept. of Applied Mathematics {\rm \&} Theoretical Physics, The Queen's University of
Belfast, Belfast, BT{\normalsize 7}  {\normalsize 1}NN, Northern Ireland }

\author{
W. T. Coffey\cite{e-cof} 
}

\address{
Department of Electronic {\rm\&} Electrical Engineering, Trinity College, Dublin 2, Ireland\\
%}
%\date{\today}
%\maketitle
%\abstract{
\smallskip
{\rm(Received 11 March 1999)}
\bigskip\\
\parbox{14.2cm}
{\rm
Classical escape rates of uniaxial spin systems are characterized by a
prefactor differing from and much smaller than that of the particle 
problem, since the
maximum of the spin energy is attained everywhere on the line of constant
latitude: $\theta=\const$, $0 \leq \varphi \leq 2\pi$.
If a transverse field is applied, a saddle point of the energy is formed, and
 high, moderate, and low damping regimes (similar to those for particles)
appear. 
Here we present the first analytical and numerical study of crossovers between 
the uniaxial and other regimes for spin systems.
It is shown that there is one HD-Uniaxial crossover, whereas at low damping the
uniaxial and LD regimes are separated by {\em two} crossovers.      
\smallskip
\begin{flushleft}
PACS numbers: 05.40.-a, 75.50.Tt
\end{flushleft}
} 
} 
\maketitle

\vspace{-2cm}
The study of thermal activation escape rates of fine magnetic particles, which 
are usually modelled as classical spins with predominantly uniaxial anisotropy, 
may be traced from the early predictions of 
N\'eel \cite{nee44}
through the first theoretical treatments of 
Brown \cite{bro63,bro79}
to the recent experiments of
Wernsdorfer {\em et al} \cite{weretal97} 
on individual magnetic particles of controlled form.
These experiments allow one for the first time to check the Stoner-Wohlfarth angular 
dependence of the switching field 
\cite{stowoh48} 
and to make a comparison \cite{cofetal98prl,cofetal98jpcm} with existing theories 
 where the energy barrier is reduced by applying a magnetic field. 
The theories checked are those for the intermediate-to-high damping (IHD) case
\cite{bro79,geocofmul97,cofetal98}, as well as for the low-damping (LD) case 
\cite{kligun90}.

The IHD and LD limits for spins are similar to those for the particle problem,
which were established by Kramers \cite{kra40}.
The most significant difference is that for spins in the HD limit the prefactor 
$\Gamma_0$ in 
the escape rate $\Gamma= \Gamma_0 \exp(-\Delta U/T)$ behaves as $\Gamma_0 \propto a$, where
 $a$ is the damping constant [if the Landau-Lifshitz (LL) equation is used], whereas for 
particles $\Gamma_0\propto 1/a$. 
A question which has not yet been addressed, both theoretically and experimentally,
and which is the subject of this Letter, is how these three damping-governed regimes merge into the single uniaxial regime
\cite{bro63} if the field is removed?

Let us consider the Hamiltonian
%\marginpar{HamTr}
%
\begin{equation}\label{HamTr}
{\cal H} = -\tilde K n_z^2  - \mu_0 {\bf n H}, \qquad |{\bf n}|=1, 
\end{equation}
where $\mu_0=M_s V$ is the magnetic moment and $\tilde K =KV$ is
the uniaxial anisotropy energy  of the particle.
The Fokker-Planck equation for the distribution function of the spins
$f(\theta,\varphi,t)$, which follows from the stochastic LL 
equation, reads
%
%\marginpar{FPESpher}
%
\begin{equation}\label{FPESpher}
\frac{ \partial f }{ \partial t } + \frac{ \partial j_x }{ \partial x }
+ \frac{ \partial j_\varphi }{ \partial \varphi } = 0,
\end{equation}
where $x\equiv \cos \theta$,
%
%\marginpar{jxphi}
%
\begin{eqnarray}\label{jxphi}
&&
j_x = \frac\gamma\mu_0 \left[
-\frac{ \partial {\cal H} }{ \partial \varphi } f - a(1-x^2)
\left( 
\frac{\partial {\cal H} }{ \partial x } + T  \frac{\partial  }{ \partial x }
\right) f \right]
\nonumber \\
&&
j_\varphi = \frac\gamma\mu_0 \left[
\frac{ \partial {\cal H} }{ \partial x } f - \frac{  a }{1-x^2 }
\left( 
\frac{\partial {\cal H} }{ \partial \varphi } + T  \frac{\partial  }{ \partial \varphi }
\right) f \right] .
\end{eqnarray}

To solve the FPE at low temperatures, $T\ll \Delta U$, we represent $f$ 
as $f({\bf N}, t) = f_{\rm eq}({\bf N}) g({\bf N}, t)$,
where $f_{\rm eq}({\bf N}) = {\cal Z}^{-1} \exp[-{\cal H}({\bf N})/T]$ is the
equilibrium distribution function.
Neglecting the exponentially small $\dot g$ one obtains
%
%\marginpar{gEq}
%
\begin{eqnarray}\label{gEq}
&&
\frac{ \partial {\cal H} }{ \partial \varphi }
\frac{ \partial g }{ \partial x }
-  \frac{ \partial {\cal H} }{ \partial x }
\frac{ \partial g }{ \partial \varphi }
+ a\left[ 
\left(
-\frac{ \partial {\cal H} }{ \partial x } + T \frac{\partial }{ \partial x } 
\right)
(1-x^2)\frac{ \partial g }{ \partial x }
\right.
\nonumber \\
&&
\left.
{} + \frac{ 1 }{1-x^2 }
\left(
-\frac{ \partial {\cal H} }{ \partial \varphi } + T \frac{\partial }{ \partial \varphi }
\right) \frac{ \partial g }{ \partial \varphi }
\right]
 = 0. 
\end{eqnarray}
The function $g$ assumes the values $g_1$ and $g_2$ in the wells and changes in
a narrow region about the top of the barrier, so that
%
%\marginpar{zetaDef}
%
\begin{equation}\label{zetaDef}
g(x,\varphi) = g_1 + (g_2-g_1) \zeta(x,\varphi),
\end{equation}
where $\zeta$ assumes the values 0 and 1 in the first and second wells,
respectively.
The numbers of particles in the wells satisfy $N_1+N_2=1$ and 
$N_i = g_i N_{i,\rm eq}$,
where $N_{i,\rm eq} = {\cal Z}_i/{\cal Z}$ are the equilibrium values and 
${\cal Z}_1$ and ${\cal Z}_2$ are partition functions for each of the wells.

The change in the particle number's in the 1st well $N_1$ is due to the flow of
particles from 1st to the 2nd well through the line $x=\const$ 
(say, the equator $x=0$):
$\dot N_1 =  -\int_0^{2\pi} d\varphi j_x$.
Using $f_{\rm eq} \partial {\cal H}/\partial \varphi = 
- T \partial f_{\rm eq}/\partial \varphi$, one can integrate the first term of 
$j_x$ by parts.
Finally, using Eq.\ (\ref{zetaDef}), one has at the kinetic
equations
%
%\marginpar{KinEq}
%
\begin{equation}\label{KinEq}
\dot N_1 = - \dot N_2 = \Gamma (N_2 N_{1, \rm eq} - N_1 N_{2, \rm eq}),
\end{equation}
 the escape rate being
%
%\marginpar{GamGen}
%
\begin{equation}\label{GamGen}
\Gamma = \left[ \frac 1 {\cal Z}_1  +\frac 1 {\cal Z}_2 \right] 
\frac{ \gamma T }{ \mu_0 }\int_0^{2\pi} \!\! d\varphi e^{-{\cal H}/T}
\left[ 
a (1-x^2)\frac{ \partial \zeta }{ \partial x } + \frac{ \partial \zeta }{ \partial \varphi } 
\right] ,
\end{equation}
where $\zeta$ satisfies Eq.\ (\ref{gEq}) with the boundary conditions stated in 
Eq.\ (\ref{zetaDef}).
Also,  one can calculate a current through 
a line of constant energy surrounding one of the wells, which is appropriate in
the LD case.

For the transverse-field (unbiased) model 
%
%\marginpar{epsDef}
%
\begin{equation}\label{epsDef}
{\cal H}(x, \varphi)/T \equiv \varepsilon =
 -\sigma( x^2 + 2h\sqrt{1-x^2} \cos \varphi),
\end{equation}
where
%
%\marginpar{RedVars}
%
\begin{equation}\label{RedVars}
\sigma \equiv \frac{ \tilde K }{ T }= \frac{ KV }{ T }, 
\qquad {\bf h} \equiv \frac{ \mu_0 {\bf H} }{ 2\tilde K }=
\frac{ M_s {\bf  H} }{ 2 K }.
\end{equation}
Another useful dimensionless variable is
$\xi  \equiv 2\sigma h = H M_s V / T$.
The function $\zeta$ satisfies
%
%\marginpar{zetaEq}
%
\begin{eqnarray}\label{zetaEq}
&&
(1-h\cos\varphi) x \left(
 a \frac{ \partial \zeta }{ \partial x } + \frac{ \partial \zeta }{ \partial \varphi }
\right)
 + h\sin\varphi \left(
 \frac{ \partial \zeta }{ \partial x } - a\frac{ \partial \zeta }{ \partial \varphi }
\right) 
\nonumber \\
&&
\qquad {} + \frac{ a }{ 2\sigma } 
\left(
 \frac{ \partial^2 \zeta }{ \partial x^2 } +\frac{ \partial^2 \zeta }{ \partial \varphi^2 }
\right) = 0, 
\end{eqnarray}
where $1-x^2 \to 1$ in the region $x\ll 1$, which is relevant in
the uniaxial and IHD cases.
The result for $\Gamma$ can be written
%
%\marginpar{GamRepr}
%
\begin{equation}\label{GamRepr}
\Gamma = \frac{ \omega_1 }{ \pi }  A \exp\left[-\frac{ \Delta U }{ T }\right], 
\qquad \omega_1 = \frac{ 2\gamma K }{ M_s }\sqrt{1-h^2},
\end{equation}
where $\Delta U={\cal H}_{\rm sad}-{\cal H}_{\rm min}$,
 $\Delta U/T = \sigma (1-h)^2$ and $\omega_1$
is the ferromagnetic resonance frequency near the bottom of the
well (the attempt frequency),
the factor $A$ describes deviations from the
transition-state theory (TST), and the lack of the factor 2 in the denominator
is due to the two symmetric wells.

In the uniaxial case, $h=0$, the function $\zeta$ is independent of $\varphi$,
and the solution of Eq.\ (\ref{zetaEq}) is
$\partial \zeta / \partial x  = \sqrt{ \sigma/\pi } \exp(-\sigma x^2)$.
Eq.\ (\ref{GamGen}) then yields \cite{bro63}
%
%\marginpar{AUniax}
%
\begin{equation}\label{AUniax}
A = 2\pi a \sqrt{ \sigma/\pi }.
\end{equation}
\begin{figure}[t]
\unitlength1cm
\begin{picture}(11,6)
\centerline{\psfig{file=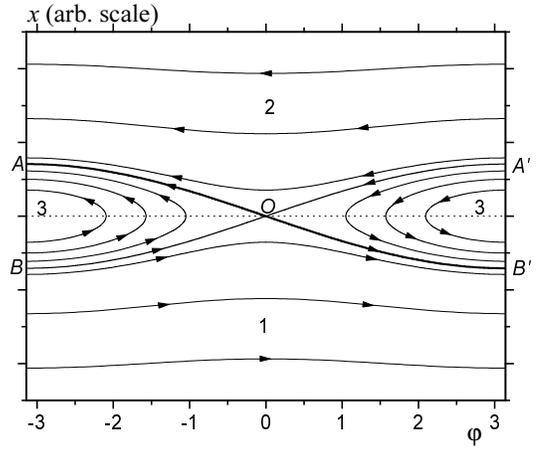,angle=-90,width=8cm}}
\end{picture}
\caption{ \label{sad_trj} 
Trajectories of a magnetic particle with uniaxial anisotropy and transverse field.
Regions 1 and 2 are the potential wells, $O$ is the saddle point, 3 is the maximum of
the potential.
}
\end{figure}

Another analytically solvable case is 
IHD with a pronounced saddle:
$T \ll {\cal H}_{\rm max}-{\cal H}_{\rm sad}$,
i.e., $\xi\equiv 2\sigma h \gg 1$ for our model.
Here the coefficients in Eq.\ (\ref{zetaEq}) for $\zeta$ can be linearized near the saddle
point $x=\varphi=0$ shown in Fig.\ \ref{sad_trj}.
Next one seeks a solution 
$\zeta(x,\varphi)=\zeta(u)$ with $u= x+\nu \varphi$ and appropriately chosen 
$\nu$ \cite{bro79,kra40}.
This form of $\zeta(x,\varphi)$ implies that it 
changes for $\sigma\gg 1$ in a narrow region across the line
$AOB'$ in Fig.\ \ref{sad_trj}.
Evaluating $\partial\zeta(u)/\partial u$, from Eq.\ (\ref{GamGen})
one obtains \cite{cofetal98}
%
%\marginpar{ASad}
%
\begin{equation}\label{ASad}
A = \frac{ a(1-2h) + \sqrt{ a^2+ 4h(1-h) } }{ 2\sqrt{h(1-h)} }
\end{equation}
in Eq.\ (\ref{GamRepr}), with the limiting forms
%
%\marginpar{ASadLims}
%
\begin{equation}\label{ASadLims}
A \cong
\left\{
\begin{array}{ll}
\displaystyle
1,           & a^2 \ll h(1-h) \\
\displaystyle
\sqrt{1-h} (1+a^2)/a,        & 1-h \ll 1, a^2 \\
\displaystyle
a/\sqrt{h},     & h \ll 1, a^2.
\end{array}
\right.
\end{equation}
Note that for $h\ll 1$ the HD regime is attained already for $a \gtrsim \sqrt{h} \ll 1$,
where the difference between the LL and Gilbert equations is still irrelevant.
For $h<1/2$, Eq.\ (\ref{ASad}) monotonically increases with $a$.
For $h>1/2$, it has a minimum at $a=2h-1$. 
In the limit $h\to 0$ the result for $A$ and thus the escape rate $\Gamma$ 
{\em diverges}, because the saddle becomes {\em infinitely wide}.
This divergence is, however, unphysical, because  
Eq.\ (\ref{ASad}) requires $\xi \equiv 2h\sigma \gg 1$.

For small $a$, the
IHD formula above fails, because the tacit assumption that
the magnetic moments approaching the barrier from the depth of the 
well are in thermal equilibrium is violated.
In the LD limit, the situation becomes completely different: The
 function $g$  changes in a narrow region along
 the  lines corresponding to the saddle-point energy $\varepsilon_c$, i.e., 
 across the lines $AOA'$ and $BOB'$ in Fig.\ \ref{sad_trj}.
Diffusion in the energy space becomes very slow, and thus  $g$
equilibrates along the  lines $\varepsilon=\const$. 
That is,  $\zeta$ of Eq.\ (\ref{zetaDef})  can be approximated as
%
%\marginpar{zetaLD}
%
\begin{equation}\label{zetaLD}
\zeta(x,\varphi) \cong \zeta(\varepsilon), \qquad  \varepsilon = \varepsilon(x,\varphi) 
\equiv {\cal H}(x,\varphi)/T.
\end{equation}
Then the conservative part of Eq.\ (\ref{gEq}) for $\zeta$
vanishes.
The rest can be averaged over the phase variable, i.e., over the constant-energy 
line, and written
%
%\marginpar{zetaLDEq}
%
\begin{equation}\label{zetaLDEq}
\frac{ d^2 \zeta }{ d \varepsilon^2 } = 
\left[ 1 - \frac{ A(\varepsilon) }{ B(\varepsilon) } \right] 
\frac{ d \zeta }{ d\varepsilon },
\end{equation}
where
%
%\marginpar{ABepsDef}
%
\begin{eqnarray}\label{ABepsDef}
&&
A(\varepsilon) \equiv \left\langle 
\frac{ \partial }{ \partial x }(1-x^2)\frac{ \partial \varepsilon }{ \partial x } 
+ \frac{ 1 }{ 1-x^2 } \frac{ \partial^2 \varepsilon }{ \partial \varphi^2 }
\right\rangle
\nonumber \\
&&
B(\varepsilon) \equiv \left \langle 
(1-x^2) \left( \frac{ \partial \varepsilon }{ \partial x }\right)^2 
+ \frac{ 1 }{ 1-x^2 } \left( \frac{ \partial \varepsilon }{ \partial \varphi }\right)^2
\right \rangle ,
\end{eqnarray}
with the averaging over conservative trajectories. 

Usually $A/B \sim T/E_{\rm char}$,
with $E_{\rm char}$ a characteristic energy, can be neglected
 at low temperatures, and Eq.\ (\ref{zetaLDEq}) is easily integrated.
Calculating the flux across the part of the separatrix 
$\varepsilon=\varepsilon_c$
around the first well yields \cite{kligun90}
%
%\marginpar{ALD}
%
\begin{equation}\label{ALD}
A= \frac{ a }{ 2 } \oint_{\varepsilon=\varepsilon_c} \left[
(1-x^2) \frac{\partial \varepsilon }{ \partial x } d\varphi
-\frac{ 1 }{ 1-x^2 } \frac{ \partial \varepsilon }{ \partial \varphi } dx
\right],
\end{equation}
i.e., $A=\tilde\delta/2$, where $\tilde\delta=\delta/T$ and $\delta$ is
the energy dissipated over the period of the precession at the saddle-point
energy in the low-damping case, coinciding with the result for particles in the LD
limit \cite{kra40}, the factor 1/2 arising because of the two symmetric wells.

If the transverse field satisfies $h \ll 1$, then Eq.\ (\ref{ALD}) greatly
simplifies.
One can set $1-x^2 \Rightarrow 1$ everywhere and retain only the leading term containing
$\partial\varepsilon/\partial x \cong  -2\sigma x$, where along the separatrix
$x = x(\varphi) \cong - \sqrt{2h(1-\cos \varphi)} = -2 \sqrt{h} \sin(\varphi/2)$.
This results in
%
%\marginpar{ALDhsmall}
%
\begin{equation}\label{ALDhsmall}
A \cong a\sigma \int_0^{2\pi} x(\varphi) d\varphi = 8a\sigma \sqrt{h},
\end{equation}
vanishing in the limit $h\to 0$, instead of approaching 
Eq.\ (\ref{AUniax}),
and becoming invalid at $\xi \lesssim 1$ (see below).

Let us now study {\em crossovers to the uniaxial regime} from the HD and LD regimes,
starting from the HD one.  
For large values of $\sigma$ one expects a crossover to the uniaxial case 
for $h \sim 1/\sigma$, which is so small that the function $\zeta$ does not yet deviate from its
uniaxial form written in the line above Eq.\ (\ref{AUniax}).
Under this assumption which will be checked shortly, the only effect of the 
transverse field is that in Eq.\ (\ref{GamGen}) 
$f_{\rm eq}\propto\exp(\xi \cos\varphi)$. 
Integration over $\varphi$ yields
%
%\marginpar{ABess}
%
\begin{equation}\label{ABess}
A = 2\pi a \sqrt{ \sigma/\pi }\exp(-\xi) I_0(\xi),
\end{equation}
which interpolates between Eq.\ (\ref{AUniax}) and the third line of 
Eq.\ (\ref{ASadLims}),
so describing a crossover between the uniaxial and HD regimes.
If $a\ll 1$, then at higher fields, $h\gtrsim a^2$, a  crossover to the 
ID regime with $A=1$ occurs, which is described by Eq.\ (\ref{ASad}).
Thus Eq.\ (\ref{ABess}) applies if
$1/\sigma \ll a^2$, i.e., $\alpha \equiv a \sqrt{\sigma}\gg 1$.

Let us now consider LD for $h\ll 1$ more
accurately.
Here $A$ and $B$ of Eq.\ (\ref{ABepsDef}) simplify to 
$A(\varepsilon) \cong \langle \partial^2 \varepsilon/\partial x^2 \rangle \cong
-2\sigma$ and 
$B(\varepsilon) \cong \langle (\partial \varepsilon/\partial x)^2 \rangle \cong
-4\sigma\varepsilon$, so that Eq.\ (\ref{zetaLDEq}) becomes
%
%\marginpar{zetaLDEq1}
%
\begin{equation}\label{zetaLDEq1}
\frac{ d^2 \zeta }{ d \varepsilon^2 } = 
\left[ 1 - \frac{ 1 }{ 2\varepsilon } \right] 
\frac{ d \zeta }{ d\varepsilon } \quad \Rightarrow \quad 
\frac{ d \zeta }{ d\varepsilon } = \frac{ C }{ \sqrt{-\varepsilon} } 
e^{\varepsilon-\varepsilon_c},
\end{equation}
where the  constant $C$ is determined by the condition
%
%\marginpar{CEq}
%
\begin{equation}\label{CEq}
2C \int_{-\infty}^{\varepsilon_c} \frac{ d\varepsilon }{ \sqrt{-\varepsilon} }
 \exp(\varepsilon-\varepsilon_c) = 1, \qquad \varepsilon_c = - \xi .
\end{equation}
Note that for $\xi \lesssim 1$ the term $1/(2\varepsilon)$ in 
Eq.\ (\ref{zetaLDEq1}) cannot be dropped, since 
$-\varepsilon \sim -\varepsilon_c = \xi$.
This is the difference from the standard LD case above.

Thus, instead of Eq.\ (\ref{ALDhsmall}),
%
%\marginpar{ALD2}
%
\begin{equation}\label{ALD2}
A \cong aC \lim_{\varepsilon\to\varepsilon_c = -\xi} 
\frac{ \oint (\partial \varepsilon/\partial x) d\varphi }{ \sqrt{-\varepsilon} }= 2\pi a \sqrt{ \sigma/\pi } f(\xi) Q,
\end{equation}
where 
%
%\marginpar{QDef}
%
\begin{equation}\label{QDef}
Q = \lim_{\varepsilon\to -\xi} \frac{ 1 }{ 2\pi }
\int_0^{2\pi} d\varphi \sqrt{ \frac{ -\varepsilon - \xi \cos\varphi }{ -\varepsilon } }
\end{equation}
and 
%
%\marginpar{fxiDef}
%
\begin{equation}\label{fxiDef}
f(\xi) = \sqrt{\pi} \left( \int_0^\infty \frac{ dt e^{-t} }{ \sqrt{\xi+t} }
\right)^{-1} =  \left( e^\xi \erfc\sqrt{\xi}\right)^{-1}.
\end{equation}
One has $Q=1$ for $\xi=0$ and $Q=2^{3/2}/\pi \approx 0.900$
for $\xi\neq 0$,
whereas the asymptotes of the function $f(\xi)$ are
%
%\marginpar{fxiLims}
%
\begin{equation}\label{fxiLims}
f(\xi) \cong
\left\{
\begin{array}{ll}
\displaystyle
1 + 2 \sqrt{\xi/\pi},        & \xi \ll 1 \\
\displaystyle
\sqrt{\pi\xi},     & \xi\gg 1.
\end{array}
\right.
\end{equation}
Thus for $\xi=0$ the uniaxial limit, Eq.\ (\ref{AUniax}), is recovered.
In the region $\xi \sim 1$, the function $f(\xi)$ describes the crossover to the
standard LD result of Eq.\ (\ref{ALDhsmall}).
The discontinuous form of $Q$ above shows that our treatment was not accurate enough to describe
yet another crossover at $\xi \sim \sqrt{\alpha}$,
where $\alpha \equiv a \sqrt{\sigma}$ is small in the LD case (see below).

It follows that the parameters governing the uniaxial crossover are
%
%\marginpar{UniCrossPars}
%
\begin{equation}\label{UniCrossPars}
\xi=2h\sigma = HM_sV/T,   \qquad    \alpha = a \sqrt{\sigma}.
\end{equation}
For $\sigma \gg 1$, in the relevant region $\xi\sim 1$ one has $h\ll 1$.
Thus in Eq.\ (\ref{zetaEq}) the terms $h\cos\varphi$,
$a\partial\zeta/\partial\varphi$, and $\partial^2\zeta/\partial\varphi^2$ can be
neglected, the derivatives with respect to $\varphi$ being much smaller than 
those with respect to $x$.
In contrast, the terms $h\sin\varphi \partial\zeta/\partial x$ and 
$\partial\zeta/\partial\varphi$ become relevant for small $a$ and should be
retained.
The resulting equation can be cast into the scaled form
%
%\marginpar{zetaEqal}
%
\begin{equation}\label{zetaEqal}
\alpha \left(
\frac{ \partial^2 \zeta }{ \partial \tilde x^2 }
+ 2\tilde x \frac{ \partial \zeta }{ \partial \tilde x }
\right) 
+ 2\tilde x \frac{ \partial \zeta }{ \partial \varphi }
+ \xi\sin\varphi \frac{ \partial \zeta }{ \partial \tilde x }=0,
\end{equation}
where $\tilde x \equiv \sqrt{\sigma} x$ and the boundary conditions are 
$\zeta=0$ for $\tilde x =-\infty$ and $\zeta=1$ for $\tilde x =\infty$ .
\begin{figure}[t]
\unitlength1cm
\begin{picture}(11,6)
\centerline{\psfig{file=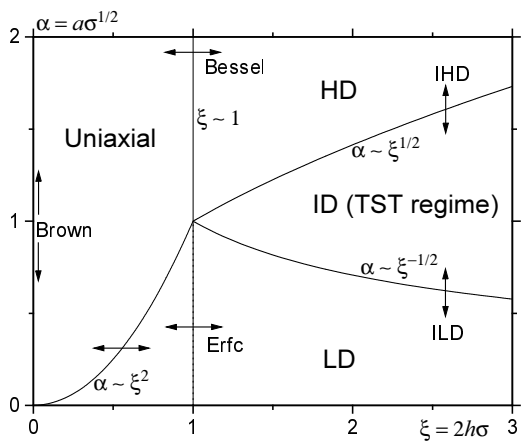,angle=-90,width=8cm}}
\end{picture}
\caption{ \label{sad_phd} 
Different regimes for the escape rate of the uniaxial spin system in the transverse field ($\sigma \gg 1$, $h\ll 1$).
}
\end{figure}

The ``phase diagram'' of the regimes for the escape rate of the
uniaxial model with transverse field is shown in Fig.\ \ref{sad_phd}. 
The ID--HD crossover is described by Eq.\ (\ref{ASad}) and occurs at $a^2 \sim h$, i.e.,
$\alpha \sim \sqrt{\xi}$.
The ID--LD crossover occurs if in Eq.\ (\ref{ALDhsmall}) $A \sim 1$, which amounts to 
$\alpha \sim 1/\sqrt{\xi}$.
This crossover is described by Melnikov's formula \cite{mel85}. 
The HD--Uniaxial crossover, which is described by 
Eq.\ (\ref{ABess}), occurs for $\alpha \gg 1$ at $\xi\sim 1$.
For $\alpha \ll 1$, there are {\em two} crossovers between the LD and uniaxial regimes.
One of them occurs at $\xi\sim 1$ and is described by Eqs.\ (\ref{ALD2}) and
(\ref{fxiDef}).
Another one occurs at $\alpha \sim \xi^2$.

The latter follows from the perturbative solution for the escape rate at small transverse
fields for arbitrary values of 
$a$.
To second order in $h$, the result has the form  
%
%\marginpar{Gamh2}
%
\begin{equation}\label{Gamh2}
\Gamma(h)/\Gamma(0) \cong 1 + (\xi^2/4) F(\alpha) = 1 + h^2\sigma^2 F(\alpha) ,
\end{equation}
where
%
%\marginpar{FRes}
%
\begin{eqnarray}\label{FRes}
&&
F(\alpha) = 1 + \frac{  1 }{\alpha^2 } \int_0^1 du \exp\left[ \frac{ \ln(1-u)+u }{ 2\alpha^2 } \right]
\nonumber\\
&&
\qquad
=1+ 2(2\alpha^2e)^{1/(2\alpha^2)} \gamma\left(1+1/(2\alpha^2), 1/(2\alpha^2)\right),
\end{eqnarray}
and $\gamma(a,z)=\int_0^z dt t^{a-1} e^{-t}$ is an incomplete gamma function.
The limiting forms of the above expression are 
%
%\marginpar{Flims}
%
\begin{equation}\label{Flims}
F \cong 
\left\{
\begin{array}{ll}
\displaystyle
1 + 1/\alpha - 1/(2\alpha)^2 + \ldots ,        & \alpha \gg 1 \\
\displaystyle
\sqrt{\pi} /\alpha - 1/3 + \sqrt{\pi} \alpha/6 + \ldots,     & \alpha \ll 1.
\end{array}
\right.
\end{equation}
The last formula shows that for $\alpha \ll 1$ the escape rate $\Gamma(h)$
essentially deviates from its uniaxial value if $\alpha \sim \xi^2$.
This defines the other crossover  mentioned above.

\begin{figure}[t]
\unitlength1cm
\begin{picture}(11,6)
\centerline{\psfig{file=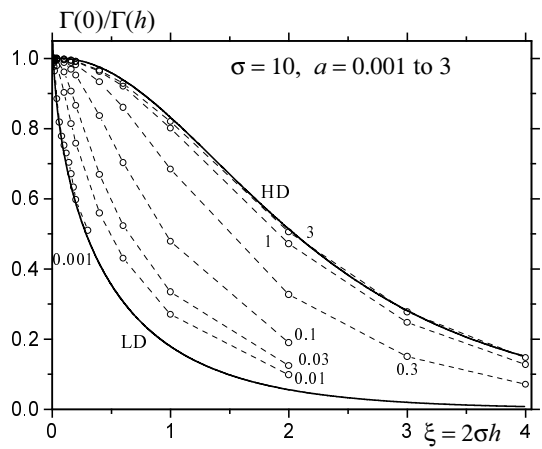,angle=-90,width=8cm}}
\end{picture}
\caption{ \label{sad_h} 
Transverse-field dependence of the inverse relaxation rate for $\sigma=10$ and different values of the
damping constant $a$.
Solid lines represent Eq.\  (\protect\ref{GamRepr}) with $A$ given by 
Eqs.\  (\protect\ref{ABess}) and (\protect\ref{ALD2}) with $Q=2^{3/2}/\pi \approx 0.900$.
}
\end{figure}

The results of the numerical calculation of the thermal activation rate as the lowest
eigenvalue of the Fokker-Planck equation \cite{cofetal98} are shown in Fig.\ \ref{sad_h}.
For large $a$ the agreement with the HD-Uniaxial crossover formula, Eq.\ (\ref{ABess}), is
rather good.
For small $a$ the field interval is restricted due to convergence problems.
Nevertheless, there good accord with the LD crossover formula, Eq.\ (\ref{ALD2}), in
the region $\xi \lesssim 1$, where the standard LD result of Eq.\ (\ref{ALDhsmall}) is
invalid.

We have shown how different damping-dependent regimes of thermal activation for uniaxial
magnetic particles with transverse field merge into the single uniaxial regime when the field tends
to zero, and presented the complete ``phase diagram'' of the different regimes.
The uniaxial characteristics appear for $\xi \lesssim 1$, i.e., for $h \lesssim 1/\sigma$  
[see Eq.\ (\ref{RedVars})].
It should be noted also that the transition from classical to quantum regimes of the escape 
with decreasing temperature is strongly modified by proceeding to the uniaxial limit 
\cite{chugar97}.
The latter is, however, a more pronounced effect and it already occurs for $h\leq 1/4$ for our 
model.

WTC thanks the Enterprise Ireland Research Collaboration fund for the support
for this work, DSFC and EK thank UKEPSRC grant nos. JR/L06225, JR/L21891.

\vspace{-0.5cm}
%\bibliography{gar}

\end{document}